\documentclass[a4paper,11pt]{article}
\usepackage{jcappub}
\usepackage{amsmath,amssymb}

\newcommand{\be}{\begin{equation}}
\newcommand{\ee}{\end{equation}}

\title{Wide Binaries and Modified Gravity (MOG)}
\author{J. W. Moffat}
\affiliation{Perimeter Institute for Theoretical Physics, Waterloo, Ontario N2L 2Y5, Canada\\
and\\
Department of Physics and Astronomy, University of Waterloo, Waterloo,\\
Ontario N2L 3G1, Canada}

\abstract{Wide binary stars are used to test the modified gravity called Scalar-Tensor-Vector Gravity or MOG. This theory is based on the additional gravitational degrees of freedom, the scalar field $G=G_N(1+\alpha)$, where $G_N$ is Newton's constant, and the massive (spin-1 graviton) vector field $\phi_\mu$. The wide binaries have separations of 2-30 kAU. The MOG acceleration law, derived from the MOG field equations and equations of motion of a massive test particle for weak gravitational fields, depends on the enhanced gravitational constant $G=G_N(1+\alpha)$ and the effective running mass $\mu$. The magnitude of $\alpha$ depends on the physical length scale or averaging scale $\ell$ of the system. The modified MOG acceleration law for weak gravitational fields predicts that for the solar system and for the wide binary star systems gravitational dynamics follows Newton's law.}

\begin{document}
\maketitle
\flushbottom

\section{Introduction}

General relativity (GR) with only ordinary baryonic matter cannot explain the present accumulation of astrophysical and cosmological data without dark matter. However, dark matter has not been observed in laboratory experiments. It is therefore important to consider a modified gravitational theory without dark matter. The difference between standard dark matter models and modified gravity is that dark matter models assume that GR is the correct theory of gravity and dark matter particles such as WIMPS, axions or fuzzy dark matter are postulated to complement the standard particle model. The MOG is described by a fully covariant action and field equations, extending GR by the addition of two gravitational degrees of freedom~\cite{Moffat2006,Moffat2021}. The first is $G = 1/\chi$, where $G$ is the coupling strength of gravity and $\chi$ is a scalar field. The second degree of freedom is a massive gravitational vector field $\phi_\mu$. The gravitational coupling of the vector field to matter is universal with the gravitational charge $Q_g=\sqrt{\alpha G_N}M$, where $\alpha$ is dimensionless, $G_N$ is Newton's gravitational constant and $M$ is the mass of a body.

We write $G=1/\chi$ as $G=G_N(1+\alpha)$, where $\alpha$ serves to measure the deviation of $G$ from $G_N$, while $\chi=1/G$ makes the action take a more conventional form. The effective running mass of the spin-1 vector graviton is determined by the parameter $\mu$, which fits galaxy rotation curves and galaxy cluster dynamics without exotic dark matter~\cite{BrownsteinMoffat2006,MoffatRahvar2013,MoffatRahvar2014,MoffatToth2014,GreenMoffat,DavariRahvar2020,MoffatToth2024}. MOG fits galaxy and galaxy cluster lensing data~\cite{MoffatTothlensing} and cosmological data~\cite{Moffat2020,DavariRahvar2021}. The merging of clusters such as the Bullet Cluster is also explained by MOG~\cite{BrownsteinMoffat,IsraelMoffat}. Analytical solutions for MOG black holes and dark compact objects have been obtained~\cite{Moffat2015a,Moffat2015b} and they have been extensively investigated in the literature.

The modified gravity model MOND proposed by Milgrom~\cite{Milgrom1983} does not fit galaxy cluster data without introducing dark matter nor does it explain cosmological data. The standard $\Lambda$CDM cosmological model has successfully fitted galaxy, galaxy cluster data as well a cosmological data with the postulate of dark matter. However, the invisible dark matter remains a mystery after decades of failure to detect a dark matter particle.

A modified acceleration law for weak gravitational fields derived from MOG, predicts that the relative velocities of wide star binaries obey Newtonian gravity, This is in contrast to the modified Newtonian gravity MOND~\cite{Milgrom1983,McGaugh2020,BanikZhou2018,BanikKroupa2019,Banik2023}, which with the MOND external field effect predicts a non-Newtonian orbital velocity of order $20\%$ greater than the Newtonian velocity.

In Section \ref{sec:FE}, we present the MOG field equations. In Section \ref{sec:motion}, we review the equations of motion of a particle and the weak field approximation of the theory, and derive the MOG weak gravitational field acceleration law and derive the Newtonian velocity prediction for wide binaries. In Section \ref{sec:tests}, we apply MOG as test of the wide binaries and discuss the MOND predicted deviation of the relative orbital binary velocity from the Newtonian velocity. In Section \ref{sec:conclusions}, we summarize our results with concluding remarks.

\section{The MOG Field Equations}
\label{sec:FE}

We formulate the MOG action and field equations in a simpler and less general way than that first published in~\cite{Moffat2006,Moffat2021}. We introduce $\chi= 1/G$ where $\chi$ is a scalar field and $G$ is the coupling strength of gravity. The MOG action is given by (we use the metric signature $(+,-,-,-)$ and units with $c=1$):
\be
S=S_G+S_\phi+S_M,
\ee
where
\be
S_G=\frac{1}{16\pi}\int d^4x\sqrt{-g}\biggl(\chi R+\frac{\omega_M}{\chi}\nabla^\mu\chi\nabla_\mu\chi 
+2\Lambda\chi\biggr),
\ee
and
\be
\label{Baction}
S_\phi=\int d^4x\sqrt{-g}\biggl(-\frac{1}{4}B^{\mu\nu}B_{\mu\nu}+\frac{1}{2}\mu^2\phi^\mu\phi_\mu-\phi_\mu J^\mu\biggr).
\ee
$S_M$ is the matter action, $\nabla_\mu$ denotes the covariant derivative with respect to the metric $g_{\mu\nu}$, $B_{\mu\nu}=\partial_\mu\phi_\nu-\partial_\nu\phi_\mu$. The Ricci tensor is
\be
R_{\mu\nu}=\partial_\lambda\Gamma^\lambda_{\mu\nu}-\partial_\nu\Gamma^\lambda_{\mu\lambda}
+\Gamma^\lambda_{\mu\nu}\Gamma^\sigma_{\lambda\sigma}-\Gamma^\sigma_{\mu\lambda}\Gamma^\lambda_{\nu\sigma}.
\ee
We expand $G$ by $G=G_N(1+\alpha)$, $\Lambda$ is the cosmological constant and $\mu$ is the effective running mass of the spin-1 graviton vector field and $\omega_M$ is a constant which we set equal to unity.

Variation of the matter action $S_M$ yields
\be
T^M_{\mu\nu}=-\frac{2}{\sqrt{-g}}\frac{\delta S_M}{\delta g^{\mu\nu}},
\ee
\be
J^\mu=-\frac{1}{\sqrt{-g}}\frac{\delta S_\phi}{\delta\phi_\mu}.
\ee
Varying the action with respect to $g_{\mu\nu}$, $\chi$ and $\phi_\mu$, we obtain the field equations:
\be
\label{Gequation}
G_{\mu\nu}=-\frac{1}{\chi^2}\biggl(\nabla_\mu\chi\nabla_\nu\chi -\frac{1}{2}g_{\mu\nu}\nabla^\alpha\chi\nabla_\alpha\chi\biggr)\\
-\frac{1}{\chi}(\nabla_\mu\chi\nabla_\nu\chi-g_{\mu\nu}\Box\chi)+\frac{8\pi}{\chi}T_{\mu\nu},
\ee
\be
\label{Bequation}
\nabla_\nu B^{\mu\nu}+\mu^2\phi^\mu=J^\mu,
\ee
\be
\label{Boxchi}
\Box\chi=\frac{8\pi}{(2\omega_M+3)}T,
\ee
where $G_{\mu\nu}=R_{\mu\nu}-\frac{1}{2}R$, $\Box=\nabla^\mu\nabla_\mu$. The energy-momentum tensor is
\be
T_{\mu\nu}=T^M_{\mu\nu}+T^\phi_{\mu\nu}+g_{\mu\nu}\frac{\chi\Lambda}{8\pi},
\ee
where
\be
T^\phi_{\mu\nu}=-\biggl({B_\mu}^\alpha B_{\alpha\nu}-\frac{1}{4}g_{\mu\nu}B^{\alpha\beta}B_{\alpha\beta}+\mu^2\phi_\mu\phi_\nu
-\frac{1}{2}g_{\mu\nu}\phi^\alpha\phi_\alpha\biggr),
\ee
and $T=g^{\mu\nu}T_{\mu\nu}$. Moreover, we have
\be
J^\mu=\kappa\rho_M u^\mu,
\ee
where $\kappa=\sqrt{G_N\alpha}$, $\rho_M$ is the density of normal, non-exotic matter and $u^\mu=dx^\mu/ds$. The vector field $\phi_\mu$ gravitational source charge is for a spherically symmetric body:
\be
Q_g(r)=\kappa M(r)=4\pi\int_0^r r'^2dr'J^0(r').
\ee
Note that if we impose the gauge condition, $\partial_\mu\phi^\mu=0$, then because of the conservation $\partial_\mu J^\mu=0$ of the current density $J^\mu$ following from \ref{Bequation} and Gauss's theorem $Q_g(r)$ is time-independent.

\section{Equations of Motion and Weak Field Approximation}
\label{sec:motion}

A massive test particle in MOG satisfies the
covariant equation of motion~\cite{Moffat2006,Roshan}:
\begin{equation}
\label{eqMotion}
m\biggl(\frac{du^\mu}{ds}+{\Gamma^\mu}_{\alpha\beta}u^\alpha
u^\beta\biggr)= q_g{B^\mu}_\nu u^\nu,
\end{equation}
where $u^\mu=dx^\mu/ds$ with $s$ the proper time along the particle
trajectory and ${\Gamma^\mu}_{\alpha\beta}$ denote the Christoffel
symbols. Moreover, $m$ and $q_g$ denote the test particle mass $m$ and
gravitational charge $q_g=\sqrt{\alpha G_N}m$, respectively.  For a
massless photon the gravitational charge vanishes,
$q_\gamma=\sqrt{\alpha G_N}m_\gamma=0$, so photons travel on null
geodesics $k^\nu\nabla_\nu k^\mu=0$~\cite{GreenMoffatToth}:
\begin{equation}
\frac{dk^\mu}{ds}+{\Gamma^\mu}_{\alpha\beta}k^\alpha k^\beta=0,
\end{equation}
where $k^\mu$ is the photon momentum and $k^2=k^\mu k_\mu=0$. This guarantees that both photons and gravitons travel on null geodesics of the theory's one metric and move with the speed of light. This is in agreement with the neutron star merger and gamma ray burster event GW170817/GRB170817A.

We note that for $q_g/m=\sqrt{\alpha G_N}$ the equation of motion for a
massive test particle (\ref{eqMotion}) {\it satisfies the (weak)
equivalence principle}, leading to the free fall of particles in a
homogeneous gravitational field, although the free-falling particles
do not follow geodesics.

In the weak field region, $r\gg 2GM$, surrounding a stationary mass
$M$ centered at $r=0$ the spherically symmetric field component $\phi_0$, with
effective mass $\mu$, can be determined by the $\phi_0$ field equation~\cite{Moffat2006}:
\be
\frac{d^2\phi_0}{dr^2}+\frac{2}{r}\frac{d\phi_0}{dr} - \mu^2\phi_0 = 0,
\ee
giving the solution:
\begin{equation}
\phi_0(r)=-Q_g\frac{\exp(-\mu r)}{r}.
\end{equation}
Here, $Q_g=\sqrt{\alpha G_N}M$ is the gravitational charge of the source mass $M$.
The radial equation of motion of a non-relativistic test particle,
with mass $m$ and at radius $r$, in the field of $M$ is then given by
\begin{equation}
\label{Eq:MOGweakEOM}
\frac{d^2r}{dt^2}+\frac{GM}{r^2}=\frac{q_gQ_g}{m}\frac{\exp(-\mu r)}{r^2}(1+\mu r).
\end{equation}
The mass $\mu$ is tiny\,---\,comparable to the experimental bound on
the mass of the photon\,---\,giving a range $\mu^{-1}$ of the
repulsive exponential term the same order of magnitude as the size of
a galaxy.  Since $ q_gQ_g/m=\alpha G_NM$, the modified Newtonian
acceleration law for a point particle can be written
as~\cite{Moffat2006,MoffatRahvar2013}:
\begin{equation}
\label{MOGaccelerationlaw}
a_{\rm MOG}(r)=-\frac{G_NM}{r^2}[1+\alpha-\alpha\exp(-\mu r)(1+\mu r)].
\end{equation}
This reduces to Newton's gravitational acceleration in the limit $\mu r\ll 1$:
\be
a_{\rm MOG}(r)=-\frac{G_NM}{r^2},
\ee
for the $\alpha$ parameter cancels. The Newtonian approximation is corrected for gravitational relativistic effects, guaranteeing that in the weak gravitational field approximation MOG can fit solar system experiments.

In the limit that $r\rightarrow\infty$, we get from
(\ref{MOGaccelerationlaw}) for approximately constant $\alpha$ and
$\mu$:
\begin{equation}
\label{AsymptoticMOG}
a_{\rm MOG}(r)\approx -\frac{G_N(1+\alpha)M}{r^2}.
\end{equation}
The MOG acceleration has a Newtonian-Kepler behavior for large $r$ with enhanced
gravitational strength $G=G_N(1+\alpha)$. The transition from Newtonian acceleration behavior for small $r$ to non-Newtonian
behavior for intermediate values of $r$ is due to the repulsive Yukawa contribution in (\ref{MOGaccelerationlaw}). This can also
result in the circular orbital rotation velocity $v_c$ having a maximum value in the transition region and account for the flatness of galaxy rotation curves without dark matter~\cite{MoffatRahvar2013}.

The MOG field equations have been solved numerically for two important cases: the spherically symmetric static vacuum solution and the cosmological case of an inhomogeneous, isotropic universe~\cite{MoffatToth2009}. These special solutions can be used in applications to fit observational data without resorting to the use of semi-phenomenological fitted parameters $\alpha$ and $\mu$. Because of the highly nonlinear nature of the coupled field equations, it is difficult to numerically solve the field equations in more general cases, although future work is planned to fulfill this task.

\section{Test of Wide Binary Gravitation}
\label{sec:tests}

The modified Newtonian dynamics MOND introduces an acceleration scale $a_0=1.2\times 10^{-10}$ m/sec$^2$. The gravitational acceleration $a$ in an isolated spherically symmetric system is asymptotically related to the Newtonian gravity $a_N$ of the baryons alone according to:
\be
a\rightarrow a_N, a_N \gg a_0\quad a_N\rightarrow \sqrt{a_0 a_N}, a_N \ll a_0.
\ee
Below the critical acceleration $a_0$ MOND deviates from Newtonian expectations~\cite{McGaugh2020}. We seek to test MOG on very small length scales by galactic standards, where the role of dark matter is greatly diminished. Since MOND does not introduce a fundamental length scale, it should remain valid in systems with much smaller masses and sizes than typical galaxies, provided that $a \ll a_0$. For a point mass $M$, this occurs beyond the MOND radius:
\be
R_{\rm MOND}=\sqrt{G_NM/a_0}.
\ee
For the Sun, this is 7000 AU (7~kAU) or 0.03 pc, much smaller than the separation of stars in the solar neighborhood. This allows for tests of MOND at kAU distances from stars in similar neighborhoods. By using stars in wide binaries with orbital accelerations less than $a_0$ and separation distances of order 7-30kAU, it is possible to test modified gravity theories. A fundamental difference between MOND and MOG is that MOG depends on the physical size and length scale of a system. This fact can be used to experimentally distinguish between MOND and MOG.

The wide binary test can be phrased as a statistical test involving the distribution of the parameter:
\be
\tilde{v}=\frac{v_{\rm rel}}{\sqrt{G_NM/r_{\rm sky}}},
\ee
where $\sqrt{G_NM/r_{\rm sky}}$ is the Newtonian circular velocity and $M$ is the total mass of the wide binary with sky-projected separation $r_{\rm sky}$ and relative velocity $v_{\rm rel}$.

Let us write (\ref{MOGaccelerationlaw}) as
\be
\gamma_{\rm grav}\equiv\frac{a_{MOG}}{a_N}=1+\alpha - \alpha\exp(-{\mu r})(1+\mu r).
\ee
We have $\mu=1/r_0$ and for $\mu r < 1$, we expand $\exp(-\mu r)$ in $\mu r$ to obtain:
\be
\gamma_{\rm grav}\sim 1+\alpha-\alpha(1-(\mu r)^2)\sim 1+\alpha(\mu r)^2.
\ee
For typical fits of MOG to galaxies without dark matter, $r_0\sim 24$ kpc. Then, we have for reasonable non-zero values of $\alpha$ and the wide binaries with $r_{\rm WB}\sim 0.1$ pc, $\mu r_{\rm WB} = r_{\rm WB}/r_0\sim 10^{-5}$ and $\gamma_{\rm grav}\sim 1$. Therefore, MOG predicts that the wide binaries satisfy Newtonian gravity.

For the fits to galaxies and cosmology data, $\alpha$ is ${\cal O}(10)$, so the parameter $\tilde{v}\leq 1$ to a high degree of confidence and MOG is Newtonian for the distribution of wide binaries. On the other hand, MOND predicts that $\tilde{v}\sim 1.5$, an $\sim 20\%$ enhancement when the external field effect is taken into account. The most detailed statistical hypothesis test to date of MOND using wide binaries with the external acceleration $a_e=1.8a_0$ reveals that Newtonian dynamics is preferred at $19\sigma$ confidence and excludes MOND at $16\sigma$ confidence~\cite{Banik2023}. Adjusting MOND interpolating functions to fit galaxy rotation curves would cause severe tension with the data. This strongly indicates that a modified gravity theory without dark matter should be dependent on the system size and length size scale \textit{not on a critical acceleration such as $a_0$}. This requirement of a system size scale dependence is fundamental to the formulation of MOG gravity. An investigation of the possible effects of an external field effect in MOG reveals that such an effect would have negligible importance~\cite{MoffatToth2021}.

\section{Conclusions}
\label{sec:conclusions}

 MOND predicts that the dynamics of a system deviates from Newtonian gravity when the acceleration $a \ll a_0$. Provided that the system has a low mass this can occur at the MOND radius $7-30\sqrt{M/M_\odot}$ kAU. This deviation ought to be detectable in wide star binaries. Analytic calculations and simulations display a $20\%$ increase in the orbital velocity over the Newtonian expectation. The covariant modified gravity theory Scalar-Tensor-Vector-Gravity or MOG predicts, on the basis of a fundamental size scale dependence, that the orbital velocity of wide binaries does not deviate from Newtonian gravity to a high degree of accuracy. A recent collaboration has used observational samples of 8611 wide binaries from Gaia DR3~\cite{Gaia2023}. A detailed statistical hypothesis test of MOND, including a calculation of the MOND gravitational field under an External Field Effect (EFE) of strength $a_e=1.8a_0$, allowed for an undetected close binary companion and possible line-of-sight contamination. A parameter $\alpha_{\rm grav}$ (not to be confused with the MOG parameter $\alpha$) is introduced, which is 0 in Newtonian gravity and MOG gravity, and 1 in MOND gravity. The statistical analysis inferred $\alpha_{\rm grav}=-0.021^{0.065}_{-0.045}$. This is consistent with Newtonian gravitation and MOG at $19\sigma$ confidence and rules out MOND at $16\sigma$ confidence. The result disagrees with the two studies~\cite{Chae2023,Hernandez2023}.
 Dark matter would not affect the wide binaries, because of their size. They are much smaller than galaxies, containing negligible amounts of dark matter, if any.

 The MOND interpolation formulas cannot be chosen to simultaneously fit galaxy and galaxy cluster data and the wide binary data without drastically affecting the galaxy constraints such as the rotation curves of galaxies. This result supports the fundamental postulate in MOG gravity that the gravitational dynamics depends on the system size scale or length dependence and not on a critical acceleration scale such as the MOND scale $a_0$. If these results obtained from the Gaia DR3 data~\cite{Banik2023} continue in future investigations to exclude MOND as a correct description of the wide binary systems, then this can mark the end of MOND as a purely acceleration dependent modification of gravity.

\acknowledgments

I thank Martin Green and Viktor Toth for helpful discussions. Research at the Perimeter Institute for Theoretical Physics is supported by the Government of Canada through industry Canada and by the Province of Ontario through the Ministry of Research and Innovation (MRI).

\end{document}